\documentclass[aps,prl,twocolumn,superscriptaddress,amsfonts,amsmath,amssymb]{revtex4-1}

\usepackage[pdftex]{graphicx}

\begin{document}

\title{Sign-reversal of the in-plane resistivity anisotropy in hole-doped iron pnictides}

\author{E.~C.~Blomberg}
\affiliation{The Ames Laboratory, Ames, Iowa 50011, USA}
\affiliation{Department of Physics and Astronomy, Iowa State University, Ames, Iowa 50011, USA }

\author{M.~A.~Tanatar}
\affiliation{The Ames Laboratory, Ames, Iowa 50011, USA}
\affiliation{Department of Physics and Astronomy, Iowa State University, Ames, Iowa 50011, USA }

\author{R.~M.~Fernandes}
\affiliation{School of Physics and Astronomy, University of Minnesota, Minneapolis, MN 55455, USA}

\author{I.~I.~Mazin}
\affiliation{Naval Research Laboratory, Code 6390, Washington, DC 20375, USA}

\author{Bing Shen}
\affiliation{Institute of Physics, Chinese Academy of Sciences, Beijing 100190, P. R. China}
\affiliation{National Laboratory of Solid State Microstructures and Department of Physics, Nanjing University, Nanjing 210093, P. R. China}

\author{Hai-Hu Wen}
\affiliation{Institute of Physics, Chinese Academy of Sciences, Beijing 100190, P. R. China}
\affiliation{National Laboratory of Solid State Microstructures and Department of Physics, Nanjing University, Nanjing 210093, P. R. China}

\author{M.~D.~Johannes}
\affiliation{Naval Research Laboratory, Code 6390, Washington, DC 20375, USA}

\author{J.~Schmalian}
\affiliation{Institute for Theory of Condensed Matter Physics and Center for Functional Nanostructutes, Karlsruhe Institute of Technology, Karlsruhe, 76131, Germany}

\author{R.~Prozorov}
\affiliation{The Ames Laboratory, Ames, Iowa 50011, USA}
\affiliation{Department of Physics and Astronomy, Iowa State University, Ames, Iowa 50011, USA }

\date{20 November 2012}

\begin{abstract}
\textbf{The concept of an electronically-driven breaking of the rotational
symmetry of a crystal \cite{Fradkin_review}, without involving magnetic
order, has found experimental support in several systems, from semiconductor
heterostructures \cite{nematic_2DEG} and ruthenates \cite{nematic_ruthenates},
to cuprate \cite{nematic_cuprates} and iron-pnictide \cite{fisherreview}
superconductors. In the pnictide BaFe$_{2}$As$_{2}$, such an ``electronic
nematic state'' appears above the magnetic transition dome, over
a temperature range that can be controlled by external strain \cite{Fscience1,detwinning}.
Here, by measuring the in-plane resistivity anisotropy, we probe the
electronic anisotropy of this material over the entire nematic/magnetic
dome, whose end points coincide with the optimal superconducting transition
temperatures. Counter-intuitively, we find that, unlike other materials,
the resistivity anisotropy in BaFe$_{2}$As$_{2}$ changes sign across
the doping phase diagram, even though the signs of the magnetic, nematic,
and orthorhombic order parameters are kept fixed. This behavior is
explained by the Fermi surface reconstruction in the magnetic phase
and spin-fluctuation scattering in the nematic phase. The unique behavior
of the transport anisotropy unveils that the primary role is played by magnetic
scattering in the normal state transport properties of the iron pnictides,
suggesting a close connection between magnetism, nematicity, and unconventional
superconductivity.}
\end{abstract}
\maketitle

\begin{figure}[htb]
\begin{centering}
\includegraphics[width=8cm]{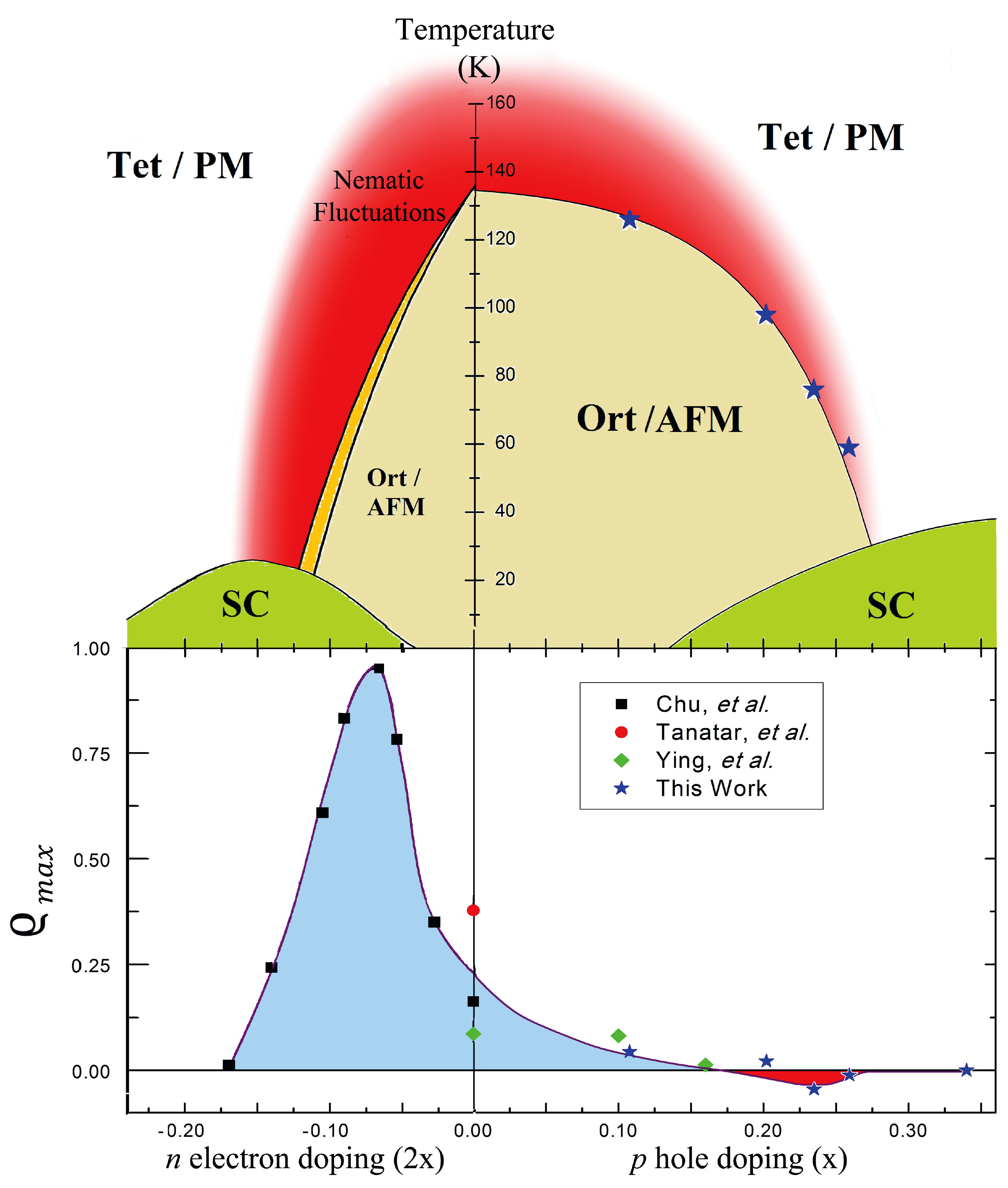}
\par\end{centering}

\caption{(Color Online) Top panel: Phase diagram of the electron- and hole-doped
Ba122 system focusing on the under - doped regime with superconducting region shown in green. The domain of the orthorhombic (antiferromagnetic) phase below transition from the tetragonal phase, $T_{S}$, is shown in yellow. Nematic fluctuations (red region) persist far above $T_{S}$ and affect the in-plane resistivity anisotropy in strained
samples. The lower panel shows the maximum in-plane anisotropy $\varrho_{\mathrm{max}}$
as a function of doping. Note the significant asymmetry of $\varrho_{\mathrm{max}}(x)$
and its sign change in the hole-doped regime.}

\label{phased}
\end{figure}

Superconductivity in all iron-based superconductors is found in the
proximity of magnetic order, which is preceded or accompanied by a
structural transition (see \cite{paglione,johnston} for review).
In the family of \emph{A}Fe$_{2}$As$_{2}$ (\emph{A}=Ca, Sr, Ba,
Eu) compounds, the magnetic order below $T_{N}$ is a uniaxial stripe
type~\cite{Cruz}, with in-plane ordering vectors $\left(\pi,0\right)$
or $\left(0,\pi\right)$. The structural transition at $T_{s}\geq T_{N}$
preempts the magnetic stripe ordering and reduces the tetragonal symmetry
of the lattice to orthorhombic. It has been suggested by several experiments
\cite{fisherreview,zxshen,Kasahara,Fscience2} and theories \cite{Kivelson,Sachdev,Mazin09,fernandes}
that the tetragonal symmetry breaking is caused by magnetic, i.e.
electronic, rather than elastic degrees of freedom, giving rise to
an electronic nematic phase, see Fig.~\ref{phased}. Doping of the
parent BaFe$_{2}$As$_{2}$ with electrons (by Co substitution of
Fe, BaCo122) or holes (by K substitution of Ba, BaK122) suppresses
both magnetic and nematic orders, revealing superconductivity with
a maximum $T_{c}$ close to the edges of the magnetic/nematic dome,
Fig.~\ref{phased}.

Below $T_{s}$, the sample splits into structural twin domains with
different directions of the orthorhombic $a_{o}$ and $b_{o}$ axes
($a_{o}>b_{o}$) \cite{domains}. Applying a small uniaxial strain
not only stabilizes one particular domain below $T_{s}$, but also
extends the nematic phase to temperatures well above\textbf{ $T_{s}$,
}due to the coupling between the strain field and nematic fluctuations
(red area in Fig.~\ref{phased}) \cite{Blombergstrain,Fscience2,shear_modulus}.
In these strained samples, the nematic state anisotropy reveals itself
not only in the small orthorhombic distortion $a_{o}-b_{o}\ll a_{o}+b_{o}$,
but, mostly, in a pronounced anisotropy of the in-plane electrical
resistivity $\varrho\equiv\rho_{b}/\rho_{a}-1$ \cite{Fscience1,detwinning,xhchen}.
Regardless of the microscopic nature of the nematic order parameter,
$\varphi$, symmetry imposes that close to the nematic transition
temperature $\varrho=\kappa\:\varphi$, where $\kappa$ is a coupling
parameter. Indeed, this relation was verified experimentally in the
parent compound of BaFe$_{2}$As$_{2}$ \cite{Blombergstrain}.
Unsurprisingly, in the only system where the electronic anisotropy was
systematically studied across the nematic dome, the ruthenate Sr$_{3}$Ru$_{2}$O$_{7}$
\cite{nematic_ruthenates}, $\varrho$ was found to retain the same
sign across the phase diagram.

In the BaFe$_{2}$As$_{2}$ family, most of the studies on its nematicity
have focused on the electron-doped compounds, finding $\varrho>0$
in both magnetic and nematic phases \cite{Fscience1,detwinning,oscillations}.
More recently, experiments in underdoped BaK122 found a rather small,
but still positive, value of $\varrho$ \cite{xhchen}. Here we report
that for larger K-doping levels, not explored in the previous studies,
the anisotropy $\varrho$ changes sign both near $T_{s}$ (nematic
phase) and at very low temperatures (magnetically ordered phase).
Since the sign of $\varphi$ is fixed by the small applied uniaxial
strain, the coupling parameter $\kappa$ must change sign across the
magnetic/nematic dome. We interpret this sign-change as a result of
asymmetric changes in the magnetic scattering and in the magnetically
reconstructed band structure for electron and hole doping. This unique
complexity of the nematic dome observed in the iron pnictides suggests
that spin-scattering dominates transport in the normal state, establishing
a possible connection to a magnetically mediated mechanism of superconductivity.

\begin{figure}[htb]
\begin{centering}
\includegraphics[width=7cm]{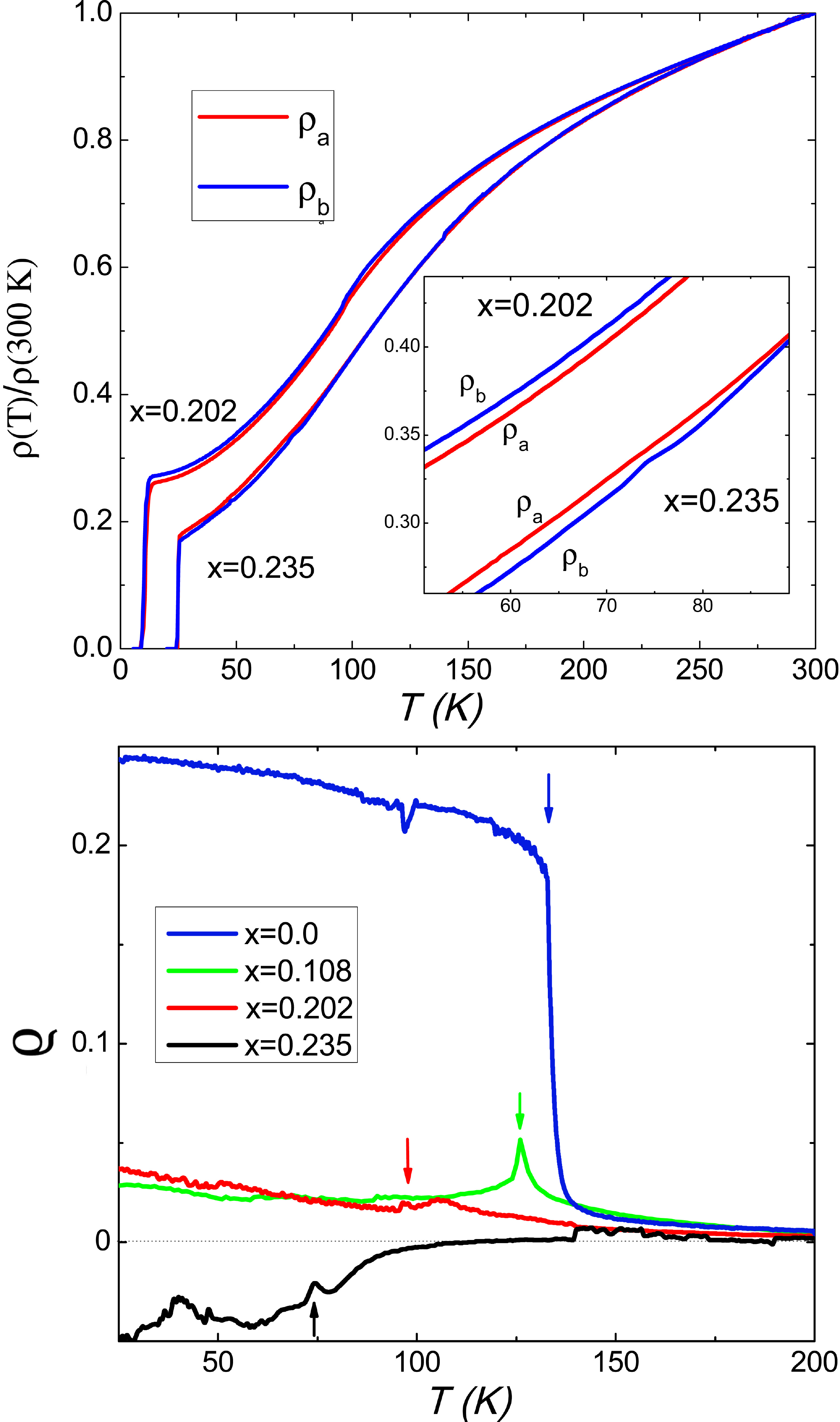}
\par\end{centering}

\caption{(Color Online) Upper panel: Normalized temperature-dependent resistivity,
$\rho(T)/\rho(300K)$, for the BaK122 samples with $x=0.202$ and
$x=0.235$. The red lines show the resistivity along the $a_{o}$-axis
($\rho_{a}$) while the blue lines show $\rho_{b}$. The inset zooms
at the structural transition, where a clear reversal of the anisotropy
from $\rho_{b}>\rho_{a}$ for $x=0.202$ to $\rho_{b}<\rho_{a}$ for
$x=0.235$ occurs. Lower panel: temperature dependence of $\varrho$
for different K-doping levels. The arrows mark the magnetic transition
temperatures.}

\label{resistivityanisotropy}
\end{figure}

Figure~\ref{resistivityanisotropy} presents our main experimental
finding, the sign reversal of the in-plane resistivity anisotropy.
For BaK122 samples with $x\leq0.202$, the resistivity is larger along
the shorter orthorhombic axis, $\rho_{a}<\rho_{b}$, whereas for samples
with $x\geq0.235$, the longer axis has higher resistivity, $\rho_{a}>\rho_{b}$.
This is clearly seen in the inset zooming in on the structural transition
region. It is also apparent that, due to the applied uniaxial strain,
the in-plane anisotropy starts to appear well above the structural
transition $T_{s}$ of the unstrained sample.

\begin{figure}[htb]
\begin{centering}
\includegraphics[width=8cm]{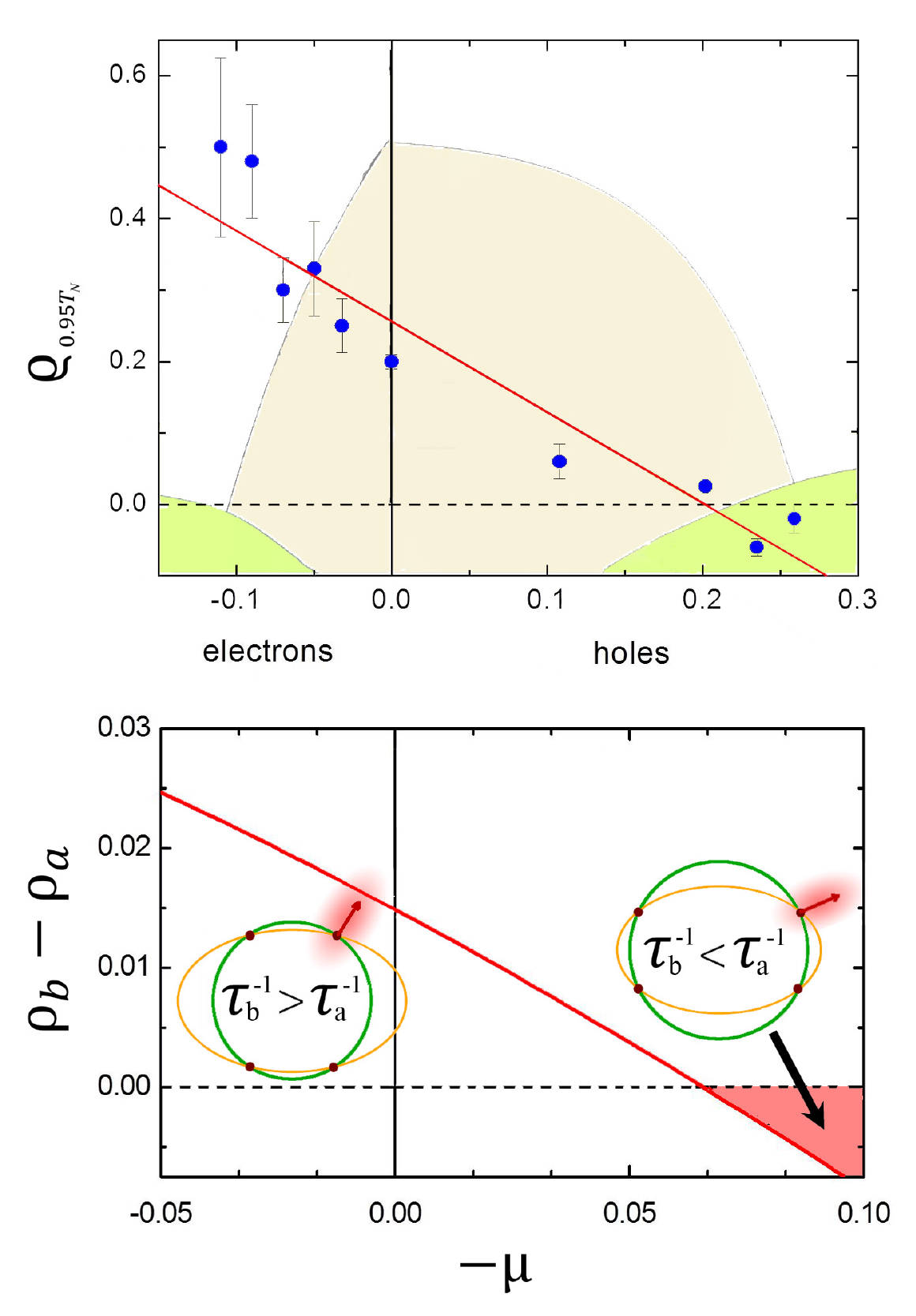}
\end{centering}
\caption{Top panel: doping dependence of the in-plane resistivity anisotropy
at $0.95T_{N}$ in both electron- and hole-doped BaFe$_{2}$As$_{2}$.
The red line is a guide to the eye. The nematic/magnetic and superconducting
domes are shown in the background for reference only, without temperature
scale. Bottom panel: theoretical calculation of the maximum in-plane
resistivity anisotropy in the paramagnetic phase normalized by the
residual resistivity, $\rho_{0}$, as a function of the chemical potential
(in units of the Fermi energy). The insets show schematically
the differences in the scattering rates corresponding to different Fermi velocities at the hot spots for electron-
and hole-doped systems. When the hot-spot Fermi velocity has a larger component along the $b$ ($a$) direction, the magnetic
scattering rate is larger along the $b$ ($a$) direction. The hot spots
are obtained by displacing the electron pocket (yellow ellipse) by
$\left(\pi,0\right)$, making it overlap with the hole pocket (green
circle). The shaded red area denotes the predicted sign reversal of
the anisotropy.}
\label{theory}
\end{figure}

In the bottom panel of Fig.~\ref{resistivityanisotropy} we plot
the temperature-dependent anisotropy ratio, $\varrho$, for several compositions.
We use two characteristic features of the $\varrho(T)$ curves
to further analyze the data. First, we plot in the bottom panel of
Fig.~\ref{phased} the maximum in-plane anisotropy $\varrho_{\mathrm{max}}$
for different $x$. In the upper panel of Fig.~\ref{theory}, we
plot $\varrho(T\approx T_{N})$, i.e. at temperatures immediately
below the magnetic transition, which coincides with the structural
transition temperature of the unstrainned BaK122 samples.

An interesting general picture emerges from the analysis of the experimental
observations of this and previous studies \cite{Fscience1,detwinning,xhchen}.
On the one hand, for all electron-doped BaCo122 and parent compositions,
the resistivity anisotropy is positive, $\varrho(x)>0$. The maximum
anisotropy, $\varrho_{\mathrm{max}}(x)$, shown in the bottom panel
of Fig.~\ref{phased}, is peaked at some intermediate electron doping,
vanishing near the edge of the magnetic/nematic dome, due to the vanishing
orthorhombic distortion. On the other hand, on the hole-doped side,
$\varrho_{\mathrm{max}}(x)$ remains positive up to moderate hole-doping,
but decreases by more than one order of magnitude, from $\varrho_{\mathrm{max}}(x=0)\approx+0.3$
to $\varrho_{\mathrm{max}}(x=0.202)\approx+0.02$. Even more surprisingly,
$\varrho_{\mathrm{max}}(x)$ changes sign and exhibits a minimum at
a higher hole doping concentration, $\varrho_{\mathrm{max}}(x=0.235)\approx-0.034$,
before it eventually approaches zero close to the edge of the magnetic/nematic
dome, near $x=0.3$.

This sign-change happens at all temperatures, even near $T_{N}$,
as shown in the bottom panel of Fig.~\ref{resistivityanisotropy}
and in the upper panel of Fig.~\ref{theory}. Furthermore, as it
can be seen in the bottom panel of Fig.~\ref{resistivityanisotropy},
the magnitude of the temperature-dependent anisotropy ratio, $\varrho(T)$,
is maximal close to $T_{N}$ in the parent and slightly hole-doped
compositions, but increases monotonically on cooling for higher hole
doping levels. This signals that the low-temperature anisotropic reconstruction
of the band structure due to long-range magnetic order plays a progressively
more important role near the hole-doped edge of the nematic/magnetic
dome.

Naively, the fact that $\rho_{b}>\rho_{a}$ across most of the phase
diagram seems surprising, because the ferromagnetic direction appears
to be less conducting than the antiferromagnetic one. From
the orbital ordering point of view \cite{velenzuela,devereaux,pp}
one expects the opposite effect. However, a closer look at the Fermi
surface reconstruction \cite{detwinning,oscillations} and the magnetic
scattering mechanisms \cite{Rafael-scattering,DagotoSF} suggests
that the anisotropy sign is decided by quantitative factors that depend
on the electronic structure, going beyond general order parameter
arguments. Moreover, the different contributions to the transport
anisotropy do not necessarily compete with each other. To sort them
out, we focus first on the nematic paramagnetic phase, in which scattering
by magnetic fluctuations dominates and\textbf{ }the band structure
is not yet reconstructed by the magnetic order.

\begin{figure}[htb]
\begin{centering}
\includegraphics[width=7cm]{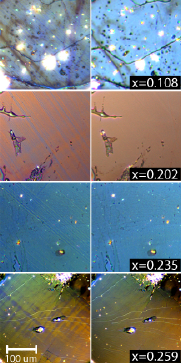}
\par\end{centering}

\caption{(Color Online) Polarized light microscopy of BaK122 samples with doping
levels as shown in the panels. Images were taken at temperatures just
above the tetragonal-to-orthorhombic structural transition (T$>$T$_{s}$,
right) and 5K (T$<$T$_{s}$, left), the latter showing formation
of structural domains due to four different orientations of $a_{O}$
and $b_{O}$ axes. The difference in color of the domains is due to
spectral dependence of bireflectance, depending on angle between the $a_{O}$
direction and the polarization plane of incident white light. Since
bireflectance is proportional to the orthorhombic distortion, the
contrast naturally vanishes in the tetragonal phase (right panels).}

\label{domains}
\end{figure}

The experimental doping evolution of $\varrho(T\approx T_{N})$, shown
in the upper panel of Fig.~\ref{theory}, displays a rather monotonic
behavior, changing sign for sufficient hole-doping levels and vanishing
at the edges of the nematic dome. Such a sign-change was previously
predicted by the theoretical model of Ref.~\cite{Rafael-scattering}
for the anisotropic magnetic scattering in the nematic phase. The
minimal model of Ref.~\cite{Rafael-scattering}, shown in the Supplementary
Material to be consistent with first-principles calculations, consists
of a circular hole pocket at the center of the square-lattice Brillouin
zone and two elliptical electron pockets centered at momenta $\left(\pi,0\right)$
and $\left(0,\pi\right)$, which coincide with the magnetic-stripe
ordering vectors. In the nematic phase, the amplitude of the fluctuations
around these two ordering vectors becomes unequal \cite{fernandes},
breaking the tetragonal symmetry of the system and inducing anisotropic
scattering. In the experiment, the applied strain selects the $\left(\pi,0\right)$
fluctuations.

The work in Ref.~\cite{Rafael-scattering} showed that, in the presence
of impurities and in the low-temperature limit - but still in the
paramagnetic phase - the scattering rates along
$a$ and $b$, and consequently the sign of the resistivity anisotropy is controlled
by the Fermi velocities of the hot spots connecting the hole and electron
pockets, see Fig.~\ref{theory}. Roughly, in electron-doped compounds,
the hot-spots Fermi velocity has a larger component along the $b$
direction, giving rise to a larger scattering rate for electrons travelling
along this direction, implying $\rho_{b}>\rho_{a}$. As the system
is doped with holes, the hole pocket expands while the electron pocket
shrinks, changing the position and the Fermi velocity of the hot spots,
which eventually acquires a larger component along the $a$ direction,
implying $\rho_{a}>\rho_{b}$. To go beyond the low-temperature paramagnetic
limit and make closer connection to the experiment, we used the same
model of Ref.~\cite{Rafael-scattering} and obtained numerically
the maximum resistivity anisotropy in the paramagnetic phase as function
of doping. The results, shown in the bottom panel of Fig.~\ref{theory},
agree qualitatively with the experimental data, capturing both the
electron-hole asymmetry of $\varrho(T\approx T_{N})$ and its sign-change
in the hole-doped regime. This suggests that magnetic scattering plays
a central role in the paramagnetic phase.

Our data in Fig.\ref{resistivityanisotropy} also show that the sign-change
in the resistivity anisotropy persists down to $T=0$,\textbf{ }deep
inside the\textbf{ }magnetic state\textbf{, }where magnetic fluctuations
are weaker. In this regime, transport should be governed\textbf{ }by
 anisotropic reconstruction of the band structure. By employing
first-principles calculations, we obtained the Fermi surface in the
magnetically ordered state and evaluated \textbf{$\varrho$} in the
relaxation-time approximation (see Supplementary Material).

Our calculations show that, in the parent compound, the reconstructed
Fermi surface contains not only quasi-isotropic hole and electron
pockets, but also Dirac cones whose crossing points are located slightly
below the Fermi level. These features are in good agreement with quantum
oscillation measurements \cite{oscillations}. Due to the large value
of their Fermi velocities, these Dirac cones dominate the transport
in the magnetic state, giving rise to an anisotropy in the resistivity
that is in agreement with the experimental observations for $x=0$.
When holes are added to the system, the chemical potential shifts
down, eventually crossing the Dirac points. Using the values extracted
by ARPES \cite{ARPES_redistribution} and quantum oscillation measurements
\cite{oscillations}, we estimate this Dirac-point crossing to happen\textbf{
}between 0.03 and 0.1 holes/Fe. For this doping concentration, the
contribution to \textbf{$\varrho$} coming from the Dirac pockets
is vanishingly small, making\textbf{ }the contribution from the remaining
pockets\textbf{ }dominant. The latter depends on details that are
beyond the accuracy of first-principles calculations, and can either
yield a small positive or negative \textbf{$\varrho$}. Yet, our first-principles
approach correctly captures the tendency of a vanishing $\varrho$
inside the magnetically ordered phase in the hole-doped samples.

\begin{figure}[htb]
\begin{centering}
\includegraphics[width=8cm]{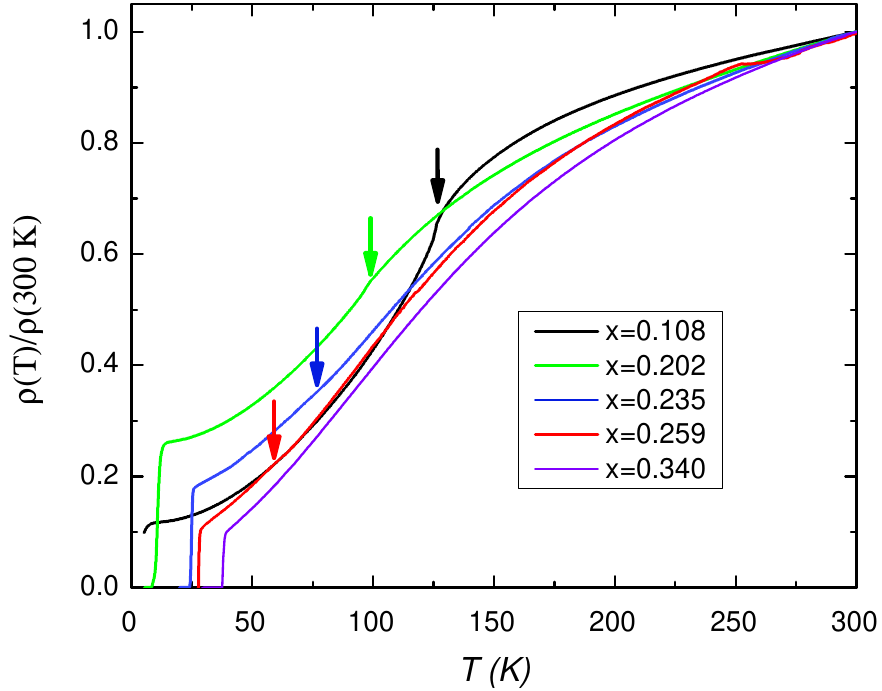}
\par\end{centering}

\caption{(Color Online) Temperature-dependent resistivity along the orthorhombic
$a_{o}$-axis, $\rho_{ao}$ for samples of BaK122 with four different
potassium doping levels in the strain-free, twinned, state. The curves
are shown using normalized plots, $\rho(T)/\rho(300K)$. Arrows indicate
the transition temperatures and were determined by the maximum in
the derivative of the resistivity and from polarized optical microscopy.
}
\label{resistivitytwinned}
\end{figure}

In conclusion, we observe the reversal of the in-plane resistivity
anisotropy between electron- and hole-doped BaFe$_{2}$As$_{2}$ compounds.
We attribute such behavior to the spin-fluctuation scattering in the
nematic phase and to the Dirac-cones contribution in the magnetically
ordered phase. Both mechanisms complementarily predict a resistivity
anisotropy that is larger in the electron-doped side and continuously
smaller in the hole-doped side. Our analysis shows that the
change in the sign of the resistivity anisotropy can be completely
understood in terms of the band structure of the pnictides, without
invoking a change from itinerant to local moment physics, as suggested
recently by Ref.~\cite{FeTe_anisotropy}. Besides shedding light on the singular
character of the nematic phase in the iron pnictides, our observation
serves as a unique fingerprint for the predominance of electronic
scattering due to spin fluctuations, rather than phonons, in the normal
state of these superconductors. Therefore, our study provides strong
support to the concept that anomalous magnetic scattering in the proximity
of a putative quantum critical point is intimately connected to magnetically-mediated
unconventional superconductivity \cite{mathur,Taillefer}.

% % % % % % % % % % % % %  SUUPLEMENTARY % % % % % % % % % % % %

\section{Supplementary Material}

\subsection{Methods}

Single crystals of BaK122 with a size up to 7x7x1 mm$^{3}$ were grown
from FeAs flux as described in Ref.\cite{s-Wencrystals}. Potassium
content in the samples was determined using electron probe microanalysis
with wavelength dispersive spectroscopy (WDS), see \cite{s-WDS} for
details. Samples, had typical dimensions of 0.5 mm wide, 2 to 3 mm
long and 0.05 mm thick, and were cut from cleaved slabs along the
tetragonal {[}110{]} direction (which becomes either the orthorhombic
$a_{o}$ or $b_{o}$ axis below $T_{SM}$). Optical imaging was performed
on the samples while mounted on a cold finger in a continuous flow
liquid helium cryostat (allowing for precise temperature control in 5K
to 300K range) using a \textit{Leica DMLM} microscope.

Polarized light microscopy was used for visualization of the structural
domains \cite{domains} and the sample selection was based on the
clarity of domains in the image.
In Fig.~\ref{domains} we show polarized light images of the area
between potential contacts in BaK122 crystals used in this study,
with $x$=0.108 (non-superconducting), $x$=0.202 (T$_{c}$=10K),
$x$=0.235 (T$_{c}$=26K) and $x$=0.259 (T$_{c}$=28K). Images were
taken on cooling at a temperature slightly above $T_{s}$ and at base
temperature. The highest contrast is observed when the sample is aligned
with the {[}100{]} tetragonal direction at a 45$^{o}$ to the polarization
direction of incident light (parallel and perpendicular to the orthorhombic
$a_{o}$ in different domains.) The contrast of domain images depends
on the quality of the surface and the homogeneity of the samples.
Domains are observed for samples with $x\approx$ 0.26, and are no
longer observed for samples with higher doping level, $x$=0.34, which
also do not show any features associated with structural transition
in $\rho(T)$.

Selected samples were mounted for
four-probe electrical resistivity measurements, with contacts made
by soldering 50 $\mu$m Ag wires using low-resistance Sn soldered
contacts \cite{s-SUST}. Initial resistivity measurements on each
sample were carried out using a flexible wire arrangement with no
strain applied to the sample (free standing state). The results are
plotted in Fig.~\ref{resistivitytwinned}. Since this measurement
is performed in the twinned state of the sample and contains contributions
from both components of the in-plane resistivity in the orthorhombic
phase, we call it $\rho_{t}$. Samples were then mounted on a brass
horseshoe straining device, and strain was applied through the voltage
contact wires by deformation of the horseshoe, see Ref.~\cite{detwinning,s-Srdetwinning}
for the details of the procedure. Strain in this configuration is
applied along the tetragonal {[}110{]} axis, which selects the orthorhombic
$a_{o}$ axis as a preferable direction upon cooling below $T_{S}$.
Very soft current leads apply no strain. The strain was incrementally
increased, and for each increment, temperature dependent resistivity
measurements were made and the domain structure was imaged to determine
the completeness of detwinning. Our previous studies using x-ray scattering
have shown that when crystals are strained to the point at which no
domains are visible in polarized microscopy, about 90 \% of the whole
bulk of the sample represents the domain whose orthorhombic $a_{O}$
axis is oriented along the direction of the strain and therefore parallel
to the current. Therefore in the detwinned state we are predominantly
measuring the $a_{o}$ axis resistivity, $\rho_{ao}$, while in the
unstrained state of the sample roughly equal contributions of both
resistivity components are measured. Therefore, we may calculate the
$b_{o}$ axis resistivity, $\rho_{bo}$(T)=2$\rho_{t}$(T) - $\rho_{ao}$(T).

\begin{figure}[htb]
\begin{centering}
\includegraphics[width=5cm]{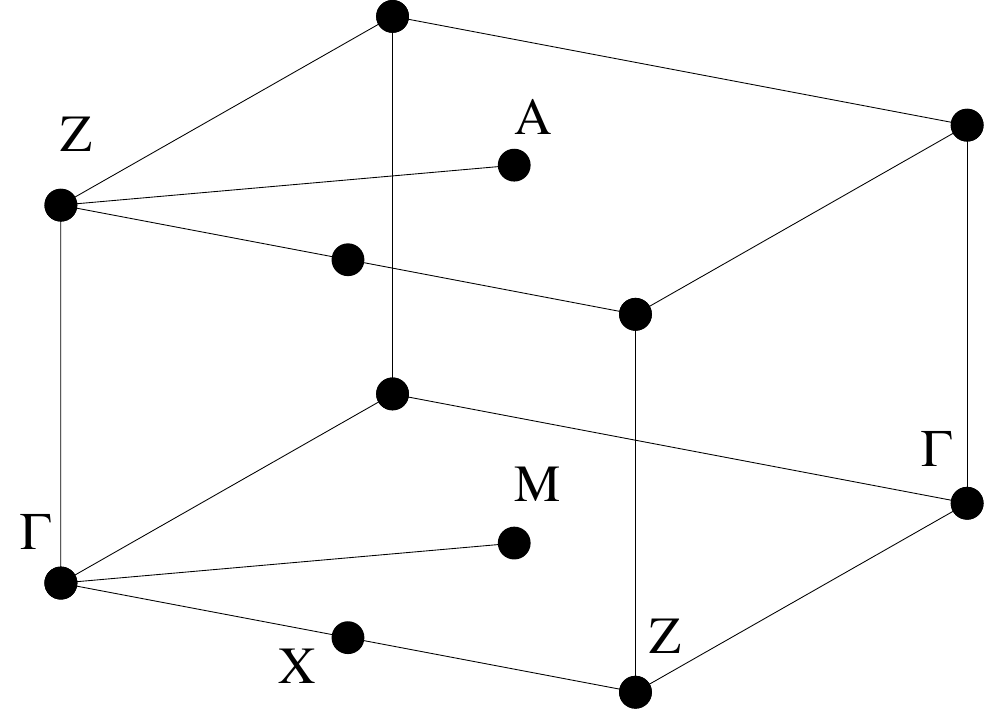}
\par\end{centering}
\caption{The Brillouin zone of BaFe$_{2}$As$_{2}$ corresponding to the two
Fe unit cell.}
\label{labels}
\end{figure}

\subsection{Contribution of the spin-fluctuation scattering}

To justify the minimal model developed in Ref. \cite{Rafael-scattering},
and used here
to calculate the resistivity anisotropy shown in Fig.~3  of the main text, we use
first-principles band structure calculations. The first step is to
unfold the first-principles Fermi surface from the two-Fe unit cell
to an effective one-Fe unit cell, where the two magnetic ordering
vectors are given by $\mathbf{Q}_{1}=\left(\pi,0\right)$ and $\mathbf{Q}_{2}=\left(0,\pi\right)$.
In the Ba122 system, due to a sizeable hybridization with the As orbitals,
there are two ways to unfold the overlapping and hybridized electron
bands. One consists of constructing Wannier functions from a particular
energy window (usually near the Fermi surface), symmetrizing them
according to the primitive tetragonal group, and recalculating the
band structure using the obtained Wannier functions \cite{s-Siggi}.
This method provides a reasonable agreement of the folded bands with
full DFT calculations, but this agreement may be achieved not through
an appropriate As-induced hybridization, but through unphysically
large one-electron hopping$.$ The other method makes use of the actual
symmetry element that reduces the larger unit cell to a smaller one:
a glide plane. This method has been discussed in details in Ref.~\cite{s-AndersenandBoeri},
and more briefly in Ref.~\cite{s-Hirschfeld-review}. It is
this unfolding procedure that is appropriate for our purposes.

\begin{figure}[htb]
\begin{centering}
\includegraphics[width=4cm]{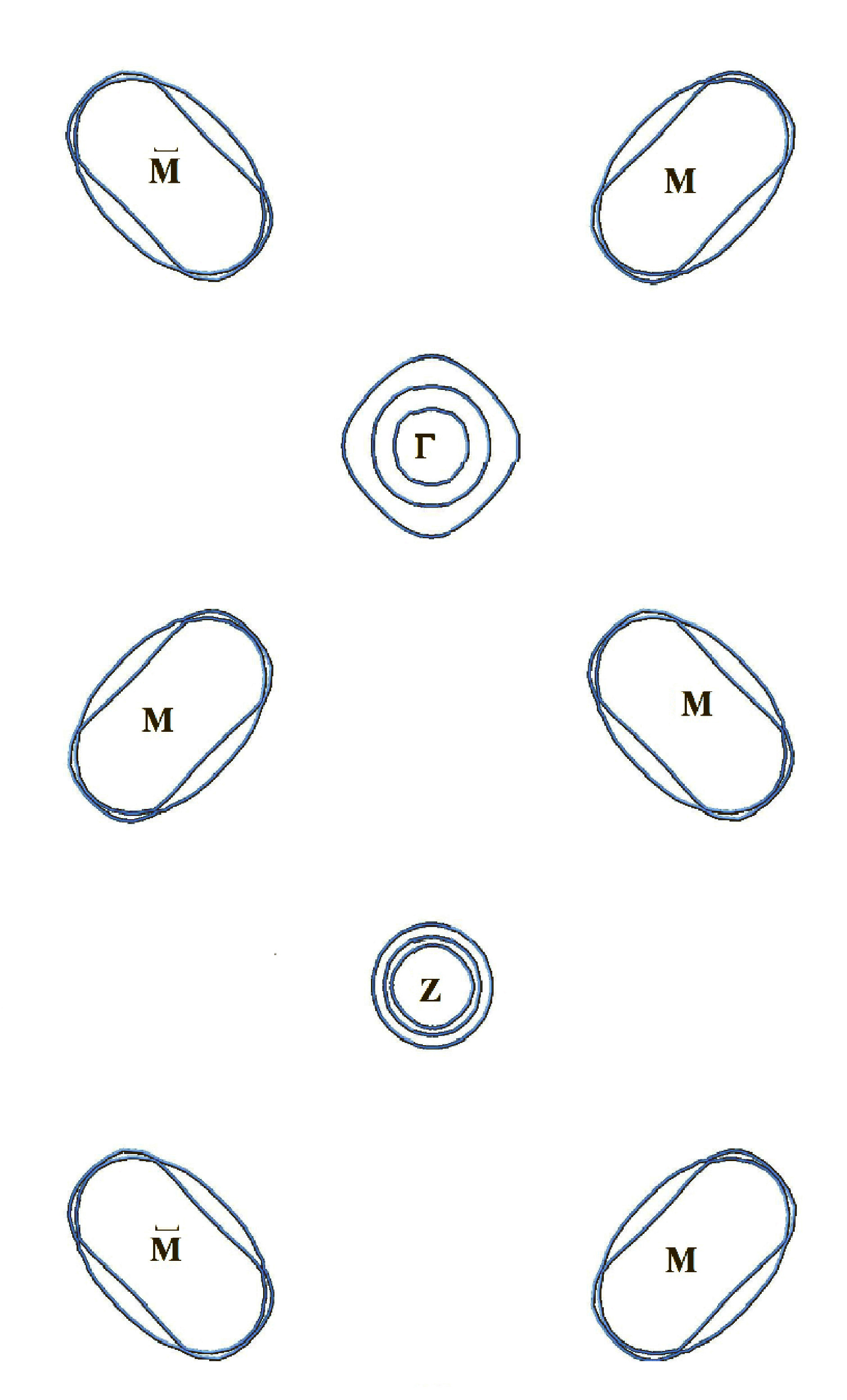}\\
\line(1,0){200}
\par
\end{centering}

\begin{centering}
\vspace{0.2cm}
 \includegraphics[width=4cm]{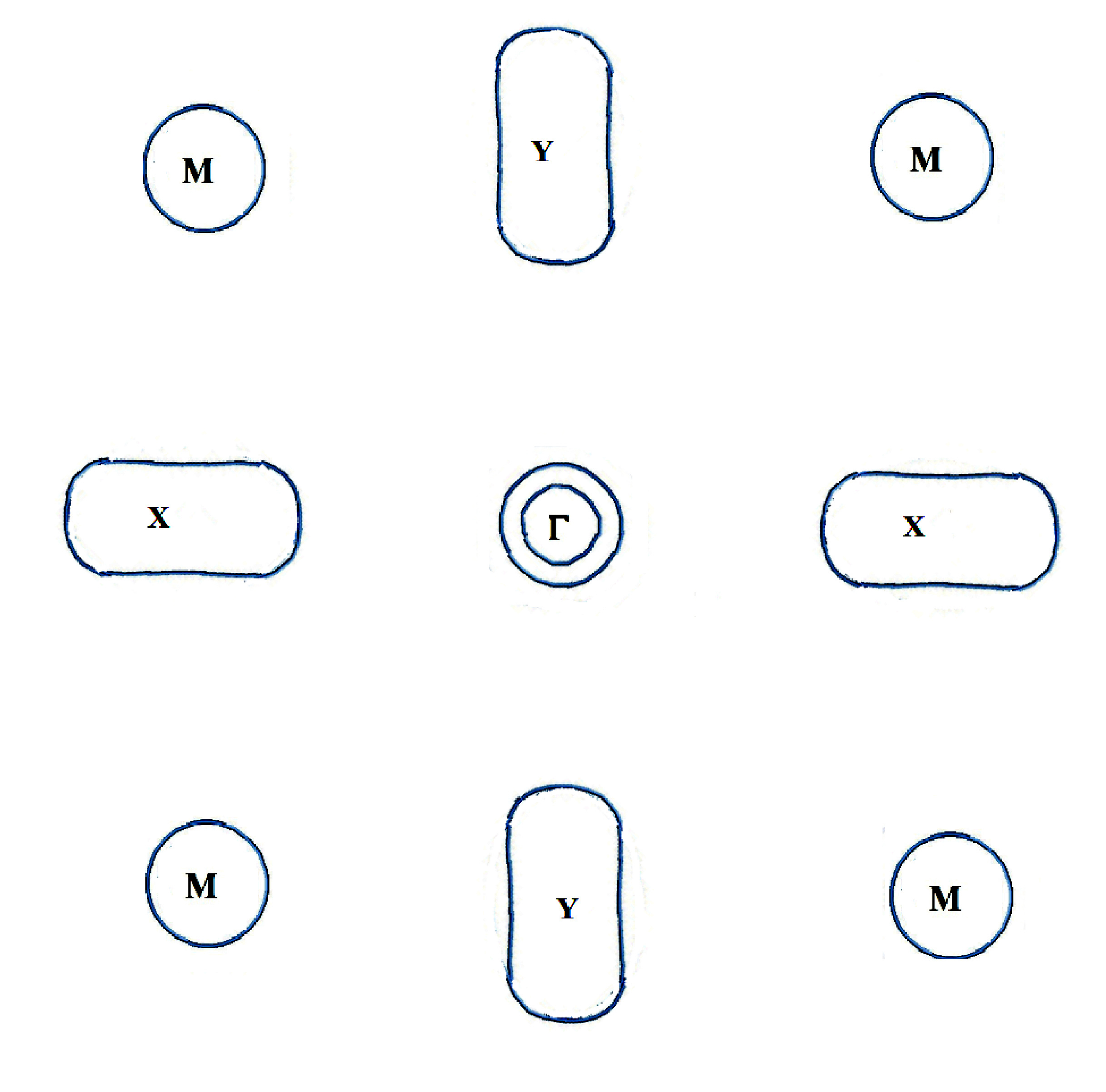}
 \line(0,1){100}
 \includegraphics[width=4cm]{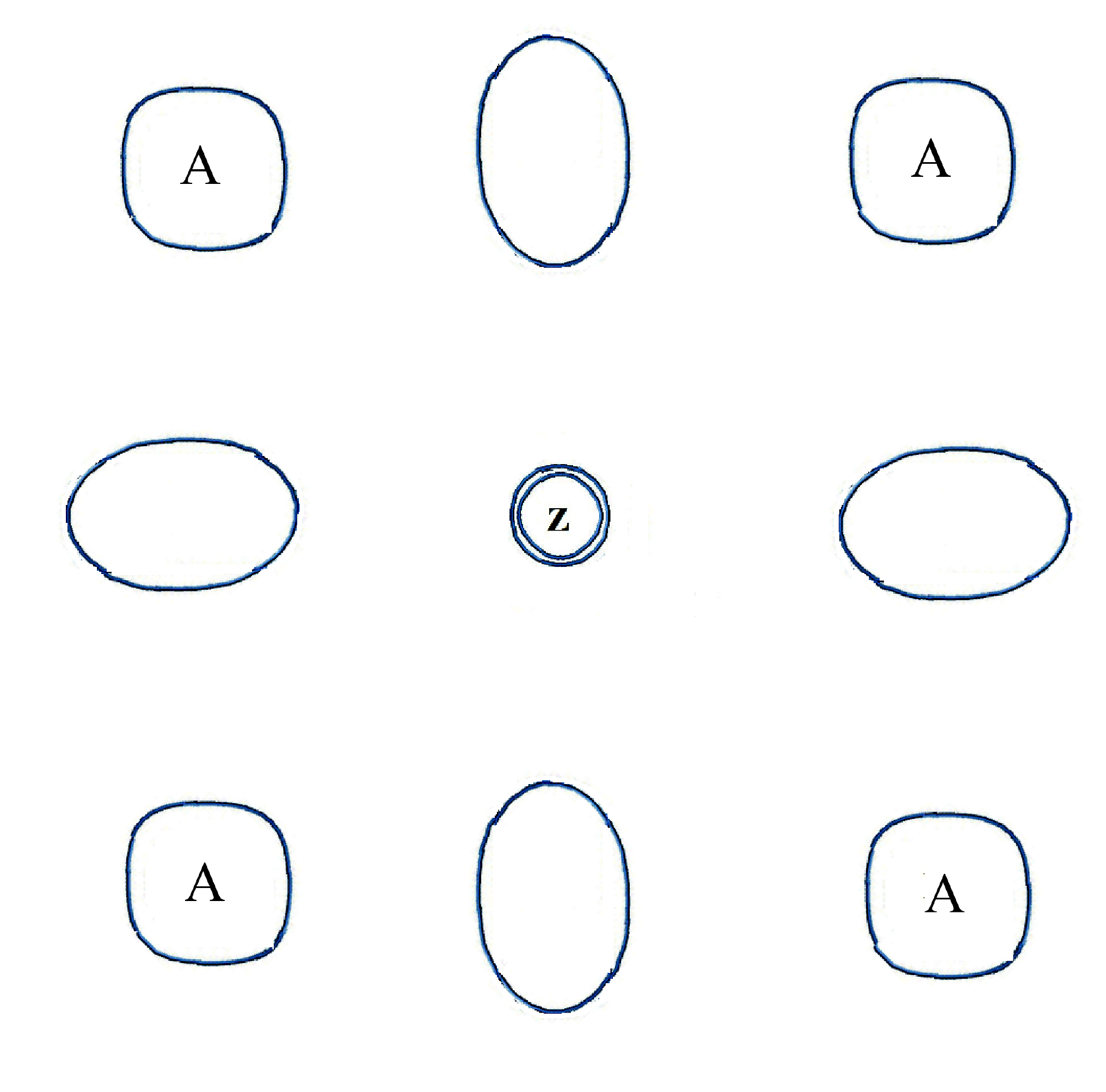}
\par\end{centering}

\caption{Upper panel: First-principles calculated Fermi surface cross-sections
for Ba122 in the folded Brillouin zone. Lower panels: the Fermi surface
obtained from the unfolding procedure discussed in the text across
$k_{z}=0$ (left panel) and $k_{z}=\pi$ (right panel). \label{fig_suppl_FS} }
\end{figure}

In the upper panel of Fig. \ref{fig_suppl_FS} we show the DFT-calculated
Fermi surface cross-sections by the plane $k_{z}=0.$ Because of the
body-centered symmetry the point Z$=(0,0,\pi)$ is equivalent to $(2\pi,0,0)$
(see Fig. \ref{labels} for notations). Therefore, the ellipticities
of the electron pockets with respect to the point Z are opposite to
those with respect to the point $\Gamma$. In the lower panel of Fig.
\ref{fig_suppl_FS} we show the Fermi surface cross-sections in the
unfolded band from which the Fermi surface of the upper panel originates.
In the unfolded zone, the ellipticity of the electron pockets with
respect to Z is the same as with respect to $\Gamma,$ but opposite
as compared to the eccentricity with respect to $M$ and $A$ (see the
notation in Fig.~\ref{labels}). After folding, the point $M$ folds
upon $Z$, and point $A$ upon $\Gamma,$ thus creating an impression
that the ellipticity changes sign with $k_{z}$.

We note that there are two hole pockets at $\Gamma$Z but only one
at $MA$, indicating that the dominating scattering processes are
between the hole pockets at $\Gamma$Z and the electron pockets. This
shows that the model of Ref. \cite{Rafael-scattering}, considering
a two-dimensional three band model with a circular hole pocket at
the center $\Gamma$ of the unfolded Brillouin zone and two elliptical
electron pockets located at $X=\left(\pi,0\right)$ and $Y=\left(0,\pi\right)$,
is an appropriate starting point. Within this model, the conductivity
tensor is given by:

\[
\sigma_{ij}=-e\sum\limits _{\mathbf{k},\lambda}v_{\mathbf{k},\lambda}^{i}f_{\mathbf{k},\lambda}^{j}
\]
 where $ $$v_{\mathbf{k},\lambda}^{i}=\partial\varepsilon_{\mathbf{k},\lambda}/\partial k_{i}$
is the component $i$ of the Fermi velocity of band $\lambda$ and
$f_{\mathbf{k},\lambda}^{j}$ is the non-equilibrium electronic distribution
function for a (unit) electric field applied along the $j$ direction.
The latter can be obtained by solving the Boltzmann equation in the
presence of scattering by impurities and spin fluctuations, as explained
in details in Ref. \cite{Rafael-scattering}. To obtain the results
of Fig.~3 , we followed the approach of Ref. \cite{Rafael-scattering},
considering a putative magnetic quantum critical point and calculating
numerically the temperature-dependent resistivity anisotropy. We used
the same parameters as those of Ref. \cite{Rafael-scattering}, varying
the chemical potential to mimic the effects of hole doping and electron
doping. For each value of the chemical potential, we obtained the
complete temperature-dependence of the paramagnetic resistivity anisotropy
and then extracted the maximum anisotropy, obtaining the results of
Fig.~3 .

\subsection{Contribution of the Dirac cones: first-principles calculations}

In order to address the kinematic effect of the magnetically-reconstructed
Fermi surface anisotropy, we calculated the transport anisotropy using
the actual DFT band structure and the constant relaxation time approximation.
It has been established that GGA calculations, even though they overstimate
the long-range ordered magnetic moment by about a factor of two, provide
a better agreement with the experimental Fermi surface than DMFT calculations,
which yield the correct magnetic moment \cite{oscillations}. This
is, incidentally, another manifestation that not only the local fluctuations,
accounted for in DMFT, but also nematic fluctuations, consistent with
a short range magnetic order of several lattice parameters, are responsible
for the reduction of the ordered moment \cite{Mazin09}. With this
in mind, we used full potential LAPW bands, with full GGA magnetization,
including spin-orbit interaction, to compute the Fermi surfaces and
the Fermi velocities.

\begin{figure}[htb]
\begin{centering}
\includegraphics[width=8cm]{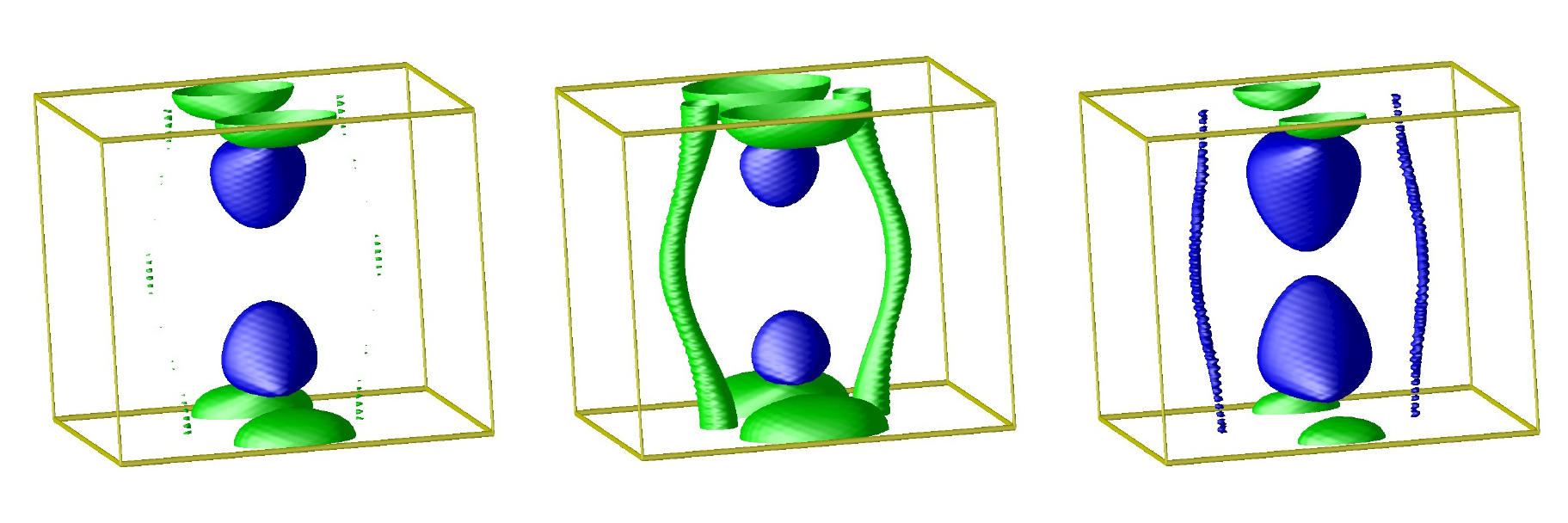}
\end{centering}

\caption{The Fermi surface calculated deep inside the magnetically ordered
phase for parent (left panel) electron- (middle panel) and hole -
doped (right panel) compositions. Green - electron pockets, blue -
hole pockets.}

\label{fig_suppl_FS2}
\end{figure}

The Fermi surface calculated deep inside the magnetically ordered
phase of the parent compound, shown in the left panel of Fig.~\ref{fig_suppl_FS2},
displays two sets of quasi-isotropic pockets: one for the holes (blue)
and one for the electrons (green). It also displays barely visible
threads, corresponding to the Dirac cones. They are green, indicating
that in the calculations the Dirac points are located very slightly
below, but practically at the Fermi level. Shifting the Fermi level
up by 30 meV transforms the Dirac pockets into clearly visible electron
tubes (middle panel in Fig.~\ref{fig_suppl_FS2}), while moving it
down (right panel in Fig.~\ref{fig_suppl_FS2}) creates hole tubes
instead. As pointed out in Ref.~\cite{s-FisherDirac}, the
Fermi velocity in the Dirac cones is high, therefore whenever they
are present, they dominate the transport. Near the Dirac point, the
contribution from a Dirac cone to transport is $Nv_{F}^{2}\propto|E|,$
where $E$ is the energy distance from the Dirac point.

\begin{figure}[htb]
\begin{centering}
\includegraphics[width=7cm]{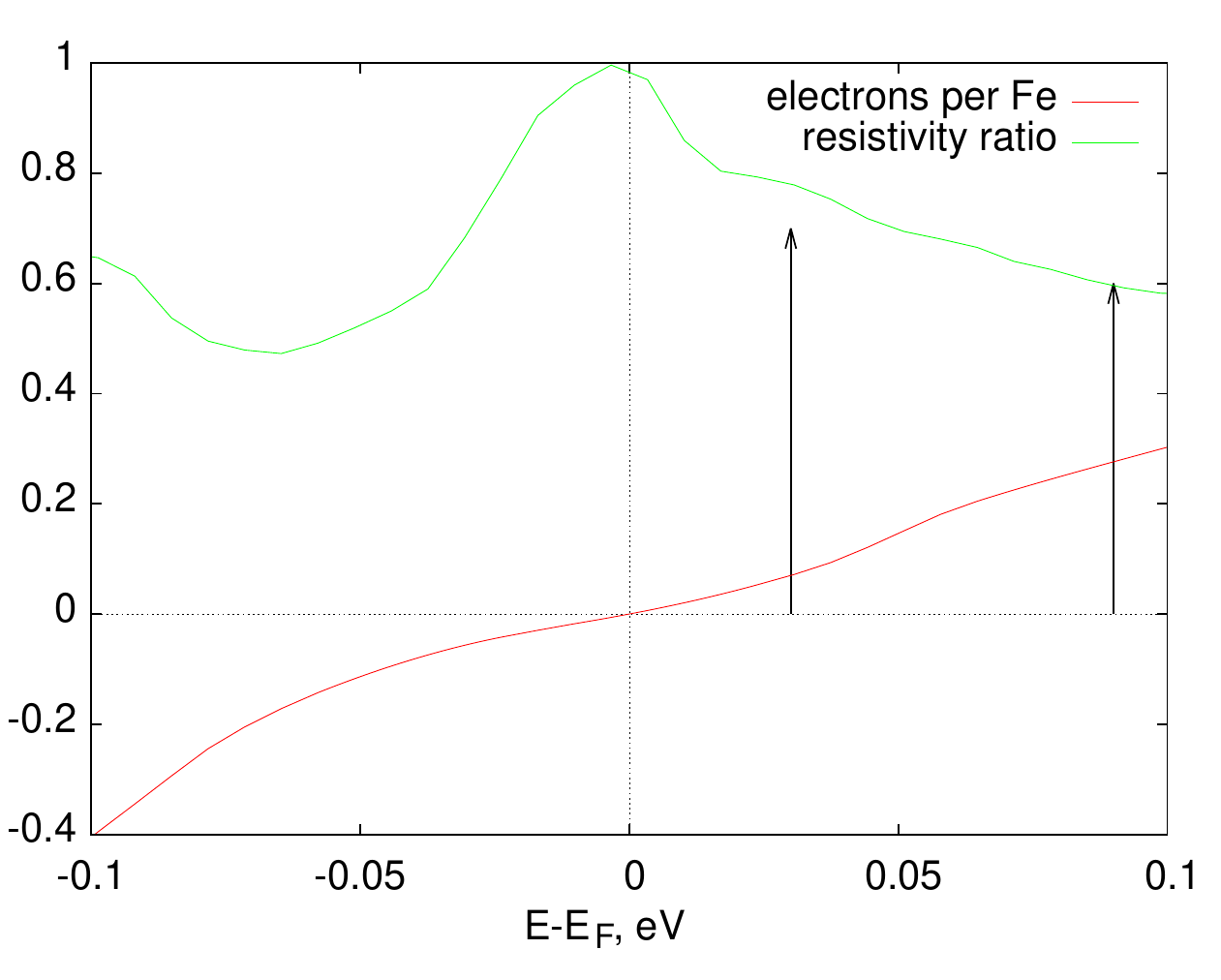}
\par\end{centering}

\caption{The hole per Fe filling factor (lower curve) and the resistivity ratio,
$\rho_{a}/\rho_{b}$ (upper curve) obtained using band structure calculation,
as described in the text. Note that the ``as-calculated'' Dirac
points lie precisely on the Fermi level, and the resistivity anisotropy
is essentially zero. In order to match the calculated Fermi surfaces
with ARPES and quantum-oscillation experiments, respectively \cite{ARPES_redistribution,oscillations},
one needs to shift the calculated Fermi level\textbf{ }of BaFe$_{2}$As$_{2}$,
correspondingly, by 30 meV and 90 meV shifts. The resulting positions
are indicated by arrows. After these shifts are applied, the minimal
anisotropy ($\rho_{a}/\rho_{b}\approx1$) appear at the left of the
arrows, that is to say, in the hole doping regime. Note that no symmetry
demands that $\rho_{a}/\rho_{b}$ be less than 1, small variation
in the calculated band structure could render it slightly larger than
1, as observed in the experiment. \label{anisotropyBS} }
\end{figure}

Using an extrafine mesh of about 60000 inequivalent k-points (44x44x40
divisions in the full Brillouin zone) we have calculated the transport
function $Nv_{F}^{2}(E)$ as a function of energy, separately for
the hole bands and electron bands. One can see very clearly two regimes:
one, where both regular and Dirac electrons contribute, and the other,
where the latter contribution disappears. Except very near the Dirac
point itself, Dirac electrons dominate transport. Moreover, contrary
to the assumption in Ref.~\cite{s-FisherDirac}, the anisotropy
comes mostly from the Dirac cones, and not from the ``round'' pockets.
As Fig.~\ref{anisotropyBS} shows, the total anisotropy is vanishingly
small when the Fermi level is crossing the Dirac cones right at the
Dirac point, which happens for \textbf{$E\approx E_{F}$ }in the calculation\textbf{.
}Away from this chemical potential, the anisotropy\textbf{ }grows,
retaining its positive value $\rho_{b}/\rho_{a}>1$, for either hole
or electron doping away from this point.

In the calculation, the Dirac point is very near the Fermi level for
the undoped compound. To determine where it is in the real material,
we can use the ARPES data of Ref. \cite{ARPES_redistribution}, or
preferably, the quantum oscillation measurements of Ref. \cite{oscillations}.
In the latter work, the authors observe that in order to bring the
Dirac cone area into a complete agreement with the experiment, they
need to shift the Fermi level in their LDA calculations up by 30 meV.
In the former work, it was observed that the Dirac points are located
30 meV below the actual Fermi level. At the same time, they observed
a factor of about three in the effective mass renormalization, implying
that, in the unrenormalized band structure, the corresponding shift
of the Fermi level would be 90 meV.

One could actually restrict the integration to particular parts of
the Brillouin zone and separate the Dirac contribution from the rest.
After that, one could shift the Dirac bands by the amount suggested
by the experiment, and the other bands by their measured shift. This
is however unnecessary, because the anisotropy is dominated by the
Dirac band, and shifts of the other bands do not really matter. Therefore,
we calculated the total anisotropy as a function of doping, using
straight DFT band structure, and simply marked (Fig.~\ref{anisotropyBS})
the position of the Fermi level of the parent compound corresponding
to the aforementioned 30 meV and 90 meV shifts. Note that the minimal
anisotropy, which corresponds to 0.03-0.1 h/Fe depending on the shift,
is nearly zero, but this is accidental; it could be a small positive
or a small negative number (as in the experiment). DFT certainly does
not describe these compounds at the level of accuracy sufficient to
distinguish between these possibilities. Yet, it consistently describes
the vanishing resistivity anisotropy in the magnetically ordered phase
as holes are doped into the system.

\section{acknowledgements}

We thank A. V. Chubukov for useful comments, suggestions and critical reading of the manuscript. RMF and JS acknowledge useful discussions with E. Abrahams. Work at The Ames Laboratory was supported by the U.S. Department of Energy, Office of Basic Energy Sciences, Division of Materials Sciences
and Engineering under contract No. DE-AC02-07CH11358. The work in China was supported by the NSF of China, the Ministry of Science and
Technology of China (973 projects: 2011CBA00102, 2012CB821403). RMF acknowledges the support of the NSF Partnerships for International
Research and Education (PIRE) program OISE-0968226.

\end{document}